# A NOVEL APPROACH TO NETWORK INTRUSION DETECTION SYSTEM USING DEEP LEARNING FOR SDN: FUTURISTIC APPROACH

## Mhmood Radhi Hadi[1] and Adnan Saher Mohammed[2]

[1]Department of Computer Engineering, Karabük University, Karabük, Turkey
[2]Adnan Saher Mohammed, Karabük University, Karabük, Turkey

## ABSTRACT

*Software-Defined Networking (SDN) is the next generation to change the architecture of traditional networks. SDN is one of the promising solutions to change the architecture of internet networks. Attacks become more common due to the centralized nature of SDN architecture. It is vital to provide security for the SDN. In this study, we propose a Network Intrusion Detection System-Deep Learning module (NIDS-DL) approach in the context of SDN. Our suggested method combines Network Intrusion Detection Systems (NIDS) with many types of deep learning algorithms. Our approach employs 12 features extracted from 41 features in the NSL-KDD dataset using a feature selection method. We employed classifiers (CNN, DNN, RNN, LSTM, and GRU). When we compare classifier scores, our technique produced accuracy results of (98.63%, 98.53%, 98.13%, 98.04%, and 97.78%) respectively. The novelty of our new approach (NIDS-DL) uses 5 deep learning classifiers and made pre-processing dataset to harvests the best results. Our proposed approach was successful in binary classification and detecting attacks, implying that our approach (NIDS-DL) might be used with great efficiency in the future.*

## 1. INTRODUCTION

The architecture of traditional networks has not changed for decades to rum that it suffers from many problems and singled out security problems. Software-defined networking new solution or approach to address these problems, and it is characterized by many features that make it the future structure of the Internet. The most prominent feature of this network is that it is inexpensive, flexible, expandable, and increases the size of its infrastructure without the complexity of the traditional network. All operations in this architecture are controlled by a controller [1]. Instructions are exchanged between the controller and the switches via the OpenFlow protocol. The SDN architecture has many advantages, as it provided many solutions to the problems of the old network infrastructure, which made it the focus of attention and interest of authors [2]. OpenFlow protocol is based on the concept of different IP packets that are exchanged between the controller and the switches. SDN provided a comprehensive overview of the entire network through the controller controlling the entire network. The controller is considered the brain of the network, which is completely isolated from the network, and targeting it from attackers means the fall of the entire network. Accordingly, the controller is the most harmful part and the most affected by attacks. It is necessary to have a network intrusion detection system (NIDS) located in the network that protects the SDN, especially the controller that is in the network part from attacks, detecting and reducing their impact. There are several types of NIDS, an approach that uses a signature, that relies on data from previous attack logs that are stored and requires continuous updating, is called the signature-based NIDS approach [3], and a second approach that uses anomaly detection that monitors the traffic pattern is more efficient and effective is called the NIDS approach Based on anomaly detection [4], which compares traffic behavior to normal and abnormal traffic. Machine learning is used with NIDS to identify attacks, but the efficiency is low. Within NIDS, a flow-based approach and anomaly detection are used together. Many factors have led to the lack of success and reliability of using machine learning in intrusion detection techniques in networks, the most prominent of which is the complexity to handle huge amounts of data that are unclassified where the performance and reliability of these systems are inefficient. Deep learning technology is a new and recent technology that predicts the possibility of solving machine learning problems, and it can deal with inconsistent data, find possible correlations, and give good and reliable performance. A reliable NIDS approach can be designed with accuracy and performance using deep learning. With deep learning, various attacks can be identified with high accuracy and with a high detection rate. SDN protection using NIDS based on deep learning is an effective method and a powerful defense mechanism. NIDS focuses on the detection of types of traffic as normal or abnormal behavior. Attacks cannot be completely prevented, but they can be detected early and identified, and their impact reduced if effective methods such as deep learning methods are used [5]. We propose a (NIDS-DL) approach for SDN using deep learning. More than one type of deep learning algorithm has been used to evaluate it based on several Metrics such as (Accuracy, F-score, Recall, Precision, etc.). we applied features selection methods to train our classifiers on high correlations features. The approach was applied to an NSL-KDD [6] dataset.

This paper is organized as follows: Section 1 Introduction. Section 2 is Related work that described some relevant previous work. Section 3 Proposed Methodology that clarified the proposed approach, also explains in brief classifiers model used and summary of architecture. Section 4 discussed the dataset and preprocessing methods applied. Section 5 Experiment results of the approach. Section 6 Study Comparative. Finally, Section 7 explains the conclusion and future work for the approach.

## 2. RELATED WORK

The application of machine learning systems with SDN has attracted the attention of many authors.

In [7] the author's purpose approach was based on five types of machine learning algorithms (RF, Naïve Bayes, SVM, CART, J84) to obtain an accurate and high-performance approach, this approach was applied to the NSL-KDD dataset with the employs 41 features, this approach achieved good detection accuracy in recognition of attacks and anomaly detection, the RF algorithm achieved the highest accuracy rate of 97%.

After the emergence of deep learning technology, several authors attempted to design several systems that use deep learning in NIDS for SDN in their approach. In [8] the authors built a deep learning-based network intrusion detection approach for the SDN environment, using the DNN algorithm in their approach. Six features from the NSL-KDD dataset used. The authors contrasted the outcomes of his approach with machine learning classifiers. The approach exhibited high detection accuracy and better performance than the machine learning classifier approach, demonstrating the feasibility and potential of using deep learning to construct network intrusion detection systems for SDN. the authors compared the results of the approach he used with machine learning classifiers.

Also, in [9] the same author proposed using a hybrid deep learning approach, the goal was to improve the accuracy and reach a better and more applicable approach, these approaches used two types of deep learning classifiers Gated recurrent unit and Recurrent Neural network to design a hybrid approach called (GRU-RNN), apply these approach was based on NSL-KDD dataset, where the author used in his approach six features in training the classifier. The hybrid approach method achieved 89% better accuracy and proved to be superior to the previous method, as well as its easy and flexible application in the SDN working environment.

Another work in [10] The goal of this approach was to build intrusion detection systems for SDN, the researcher used machine learning and deep learning systems to compare the results. A deep learning algorithm (GRU) was used in the approach, the algorithm achieved better accuracy and performance than machine learning classifiers, more than one type of dataset was used in training and comparison, six types of different attacks were categorized with a benign approach, the approach achieved great success indicating the possibility of applying deep learning in NIDS with great efficiency to SDN.

In this paper, several types of deep learning classifiers (CNN, DNN, RNN, LSTM, GRU) are applied. NSL-KDD dataset was used as the approach was applied to 12 features extracted. Each classifier was evaluated based on a different set of metrics. A broad approach to deep learning and its classifiers has been used to build a robust and effective NIDS system in detection and identifying attacks for future application within the SDN environment, which differs from the rest of the research in that it relies on more than metrics in assessment, not just accuracy and trying to get the best and highest result compared to related work.

## 3. Proposed Methodology

### 3.1. System Methodology Description

The adoption of most of the methods applied in the machine learning approach will become less effective with the development of attack and penetration systems and the tools used for them. Machine learning method needs more configured data and it also needs less data to process, moreover performance and accuracy become poor. Most of the methods that use deep learning, discussed by the authors, use classifiers. The classifier is mainly evaluated on the accuracy of the matching metric, and the accuracy is also low, which does not lead to building a reliable and efficient NIDS system to detect attacks.

All of these prompted us to build our methodology shown in Figure 1, this methodology is based on building the NIDS-DL approach for SDN, this approach uses more than one classifier for deep learning with training classifiers on 12 features extracted from 41 features in the NSL-KDD dataset, training the classifier on best correlation features will lead to the possibility of detecting various attacks. Applied feature selection method to select the best features that are effective and get correlations on the result, also the system will be powerful and reliable against attacks. The approach is evaluated on several Metrics and the classifiers are compared with each other.

In our approach, we evaluated CNN, DNN, RNN, LSTM, GRU classifiers are used, Results are compared where the (normalization) mechanism is used on the data to speed up the training process and get the best possible outcomes for generating an efficient NIDS classifier, also using feature selection method to avoid missing in training algorithm and try to reach the best accuracy and performance through selecting the best feature for training.

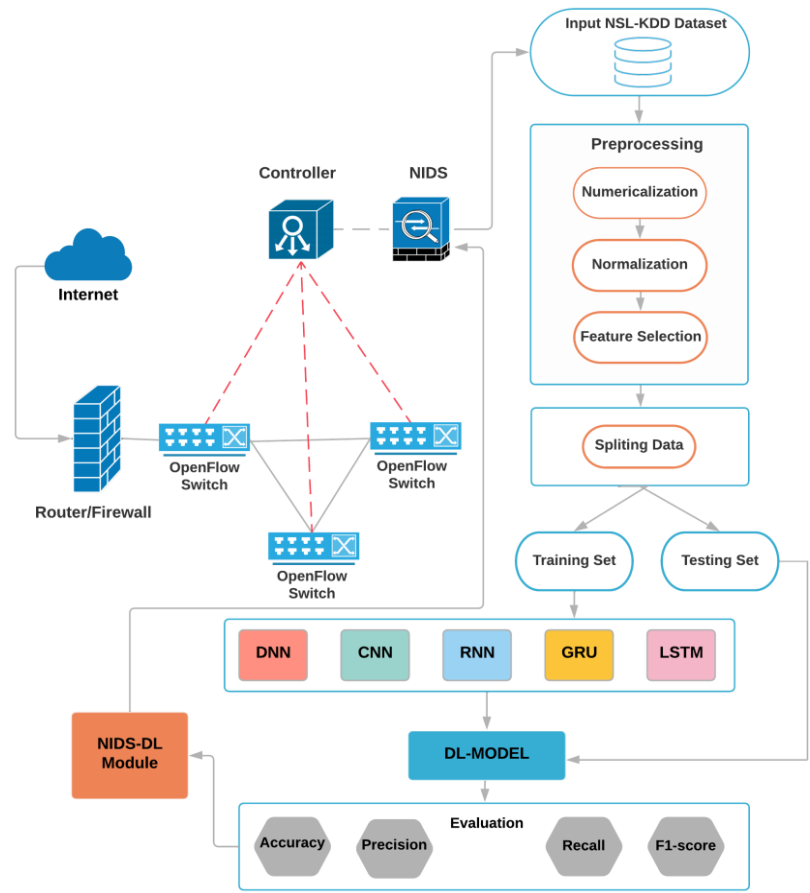

Figure 1.  Proposed Methodology for (NIDS-DL) in SDN

## 3.2. Model Classifiers

In our study we use DNN, CNN, RNN, LSTM, GRU classifiers, architectures summary is given visualization in Figures 2-6.

a)  DNN is a deep neural network is a developed class of a simple neural network. a deep neural network is called when it consists of more than three hidden layers. Increase the number of hidden layers, will be led to need for additional computer resources for processing, also that will raise ability and efficiency to process a large amount of data.

b)  CNN is a convolutional neural network that processes and classifies input in the form of images. This type of neural network has the property of extracting information and reducing features and this is reason makes it widely used in most applications. CNN uses a feed-forward feature when processing.

c)  RNN is Recurrent neural networks are also considered one of the simple neural networks, also considered a powerful type developed in the eighties. The most important thing that distinguishes this type and makes it a strong type is that it contains the internal memory

d)　LSTM is Long short-term memory is one of the types of a type of RNN. It came to address several problems that the RNN suffers from. LSTM has the feature of retaining data and information stored for a long period.

e)　GRU is Gated Recurrent Unit is also a type of standard recursive network. The specific architecture and interior design are similar to LSTM. Gated Recurrent Unit is designed to address the vanishing gradient problem in RNN.

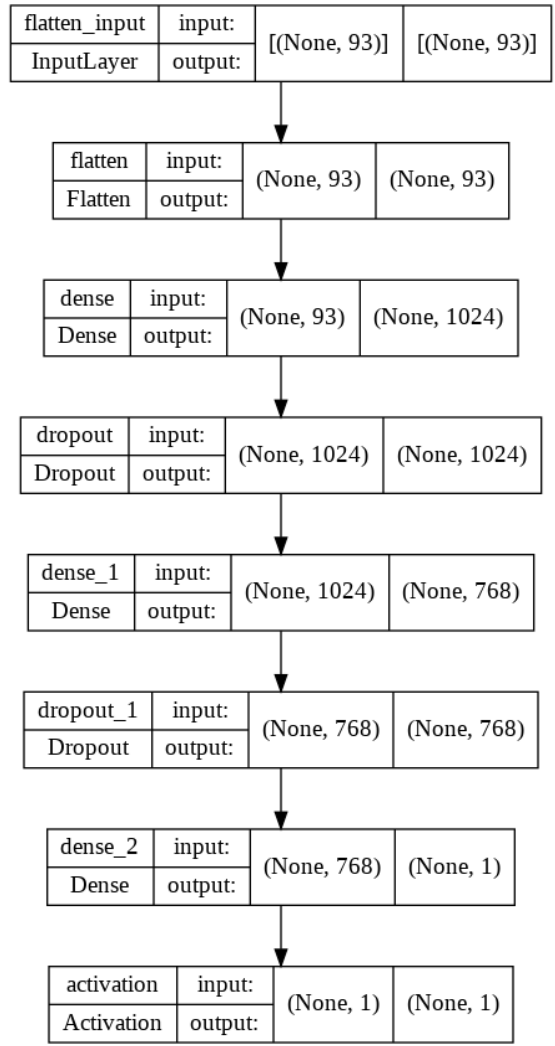

Figure 2.　Summary of DNN model.

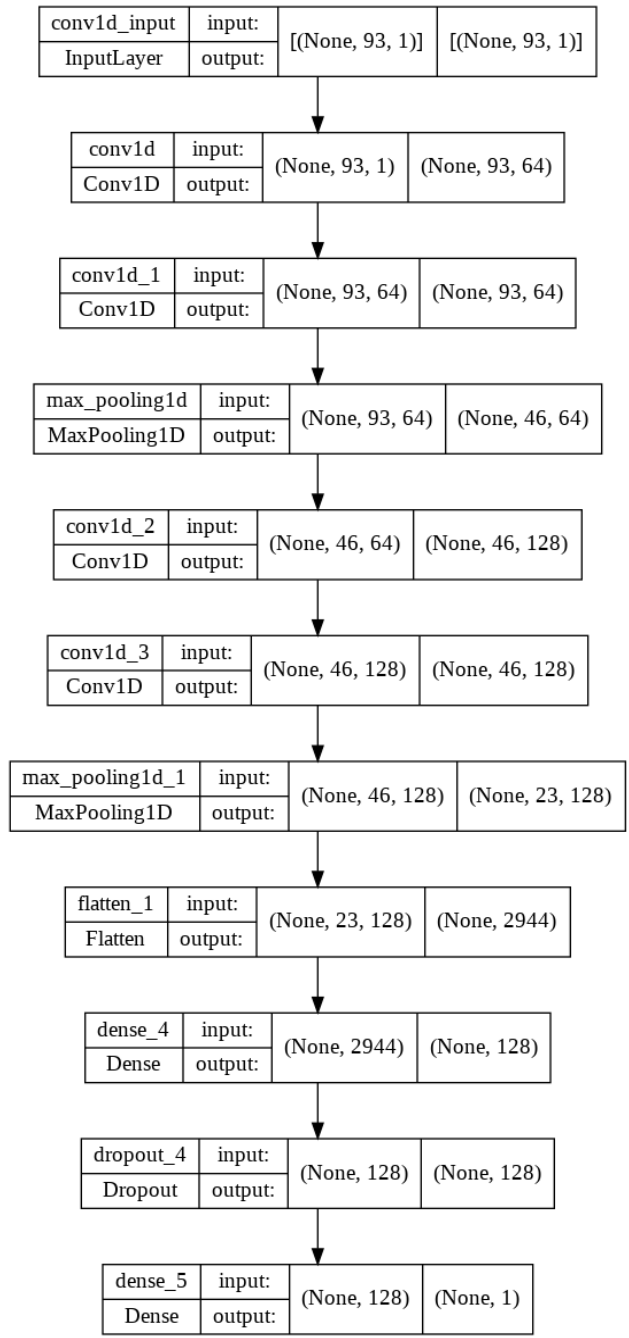

Figure 3.  Summary of CNN model.

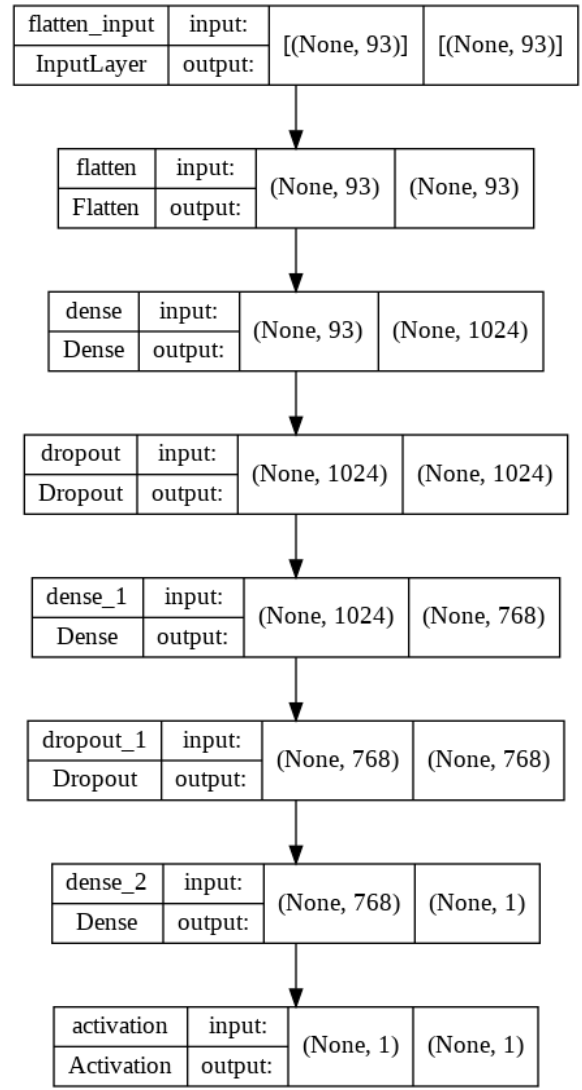

Figure 4. Summary of RNN model.

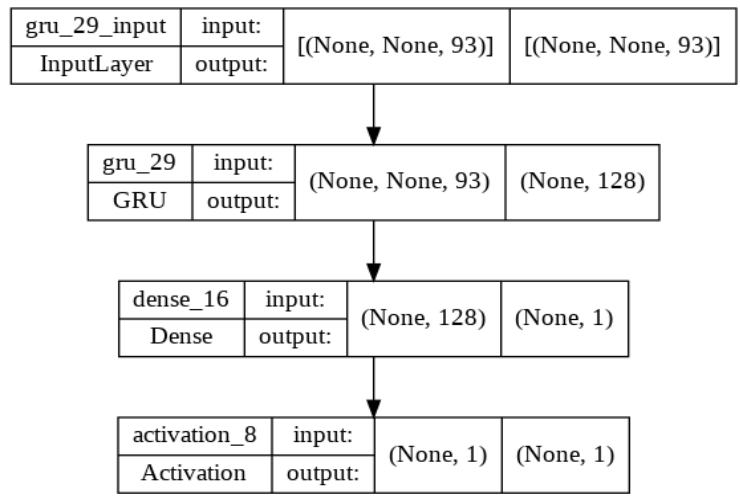

Figure 5. Summary of GRU model.

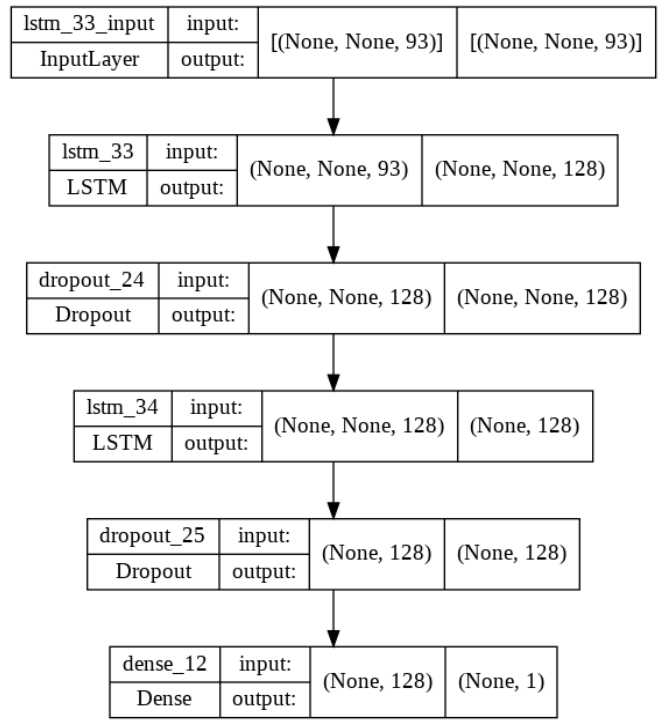

Figure 6.  Summary of LSTM model.

## 4. DATASET

In this part, we will discuss the NSL-KDD [11] dataset that was used in our proposed approach. The NSL-KDD dataset is an update and development of the KDDCup99 dataset [12], which is much older than it was proposed in 1999, as it contained several problems and contained null or it is a recursive dataset which many of its problems have been solved in the NSL-KDD dataset, but this does not mean that it does not contain mistakes. NSL-KDD contains 41 features, we extracted 12 features are more corrections using the feature selection method. NSL-KDD is used as a simulator for network data and internet traffic as it was used in several research and applied by authors in their approach. The main feature of the NSL-KDD dataset that made it preferable to many authors is that its size is almost consistent and contains reasonable several features that help in obtaining the best and most reliable classifiers.

### 4.1. Data Preprocessing

In this section, we will discuss the methods used in preprocessing datasets.

#### 4.1.1.  Numericalization

To handle the NSL-KDD dataset into deep learning classifiers, all data must be in numeric format. The NSL-KDD dataset contains three non-numeric features and 38 numeric features. The features are converted to numeric form so that they can be handled by classifiers after they are converted to array form. The features that are converted are ('flag', 'service', 'protocol_type'). For example, the feature ('protocol_type') contains three types of data ('icmp', 'udp','tcp'), which are encoded into (1,0,1), (1,1,0), (0,0,1). After using this method, all the 12 turns into a map of 122-dimensions.

**4.1.2.  Normalization**

The normalization mechanism is applied for several tasks, it is used to speed up the training process for classifiers as it works to make the data set consistent and make the difference between the data small when we have the difference between the big and small data is large. Among the features in the NSL-KDD data set and contains the difference between its data are dst_bytes [0,9.11×109], duration [0,58329], src_bytes [0,9.11×109]. The formula shown in 1 is applied, which transforms the data range and makes it between [0,1].

$$xi = (xi - Min) / (Max - Min) \qquad (1)$$

**4.1.3.  Feature Selection**

In this processing method, we extracted the features that are most correlated to the target feature, and the purpose is to reduce the loss of the classifier during training and try to get the best accurate results and high performance. Table 1. illustrates 12 features extracted from the NSL-KDD dataset.

Table 1.  Feature extracted from NSL-KDD dataset.

| No. | Features | No. | Features |
|---|---|---|---|
| 1 | protocol_type | 7 | srv_serror_rate |
| 2 | service | 8 | same_srv_rate |
| 3 | flag | 9 | dst_host_srv_count |
| 4 | count | 10 | dst_host_same_srv_rate |
| 5 | logged_in | 11 | dst_host_serror_rate |
| 6 | serror_rate | 12 | dst_host_srv_serror_rate |

**4.1.4.  Data Splitting**

The features are a selection from NSL-KDD Dataset are splitting by 75% for training and 25% for testing. Table 2. Showing partitioning of training and testing data into the NSL-KDD dataset with 12 features.

Table 2.  A distribution instance of the NSL KDD dataset.

| | Training set | Test set |
|---|---|---|
| Number of instances | 107,077 | 18,896 |

**4.2. Evaluation Metrics**

NIDS performance is evaluated by several different metrics, the most prominent of which are Accuracy (AC), Precision (P), recall (R), and F1-score (F). These metrics must be of the highest value, especially the accuracy on which NIDS reliability depends. Another is centered which is the confusion matrix within which several parameters are calculated. One of these parameters is True Positive (TP), which indicates the number of attacks that are successfully categorized as attacks. True Negative (TN) represents the number of ratings of normal records that are correctly categorized as normal. False Positive (FP) refers to the number of normal records that are incorrectly classified as attack records. False Negative (FN) indicates the number of records for attacks that are incorrectly categorized as normal records.

- Accuracy (AC): Calculate the total number of true classifications.

$$AC = (TP+TN) / (TP+TN+FP+FN) \qquad (2)$$

- Precision (P): It calculates the true classifications that NIDS can predict.

$$P = TP / (TP+FP) \qquad (3)$$

- Recall (R): It calculates the number of correct classifications compared to each intrusion.

$$R= TP / (TP+FN) \qquad (4)$$

- F1-score (F1): It is a method for calculating the harmonic mean of precision and recall.

$$F1= 2×precision×recall / precision ×recall \qquad (5)$$

## 5. Implementation of Models

Our classifiers are trained in the Google Collab environment using the Keras and Scikit-Learn libraries. The DNN, CNN, RNN, LSTM, GRU models were trained with 100 epochs and using the Adam optimizer with a learning rate of 0.01 for all classifiers. The loss type for all classifiers is also binary cross entropy with a validation distribution of 0.2.

## 6. Experiment Results

The goal of our approach is to try to get the best results for several metrics. The approach was made and implemented using the Python 3.5.6 programming language, also using (TensorFlow, Keras) with (NumPy, Pandas) library for preprocessing. The computer Hardware configuration is (Intel i7-2720 QM, 16 GB of RAM, AMD Radeon 2 GB, 256 GB SSD).

The algorithm results are presented for all algorithms in our approach using the metrics in (Accuracy, Precision, Recall, F1 score). The CNN classifier performed better than the other classifiers used in the metric (Accuracy, Precision, F1-score), with results (0.9863, 0.9845, 0.9872), respectively. The DNN classifier showed good results and was ranked after the CNN classifier by metrics (Accuracy, Precision, F1 score) and the results were (0.9853, 0.983, 0.9863) or better than these results. The rest of the classifiers except CNN. The RNN classifier obtained the best result in terms of metric (Recall) with (0.9902), outperforming all classifiers. The results of the LSTM algorithm are metrically similar (Recall) to GRU, in that it also obtains results with the metric (Accuracy, Precision, Recall, F1-score) giving the corresponding results (0.9804, 0.9767, 0.9856, 0.9816). The GRU classifier generated the (accuracy, precision, recall, F1 score) scores (0.9813, 0.98, 0.98, 0.982) respectively, resulting in the lowest score compared to the other classifiers. The GRU classifier gets close results and it looks like a valuable result, but it is low compared to the other classifiers shown in Figure 7 and Table 3.

Table 3.  Evaluation Metrics Classifiers.

| DL-Algorithm | Accuracy | Precision | Recall | F1-score |
|---|---|---|---|---|
| DNN | 0.9853 | 0.983 | 0.9896 | 0.9863 |
| CNN | 0.9863 | 0.9845 | 0.9898 | 0.9872 |
| RNN | 0.9813 | 0.9751 | 0.9902 | 0.9826 |
| LSTM | 0.9804 | 0.9767 | 0.9856 | 0.9816 |
| GRU | 0.9778 | 0.973 | 0.9856 | 0.9793 |

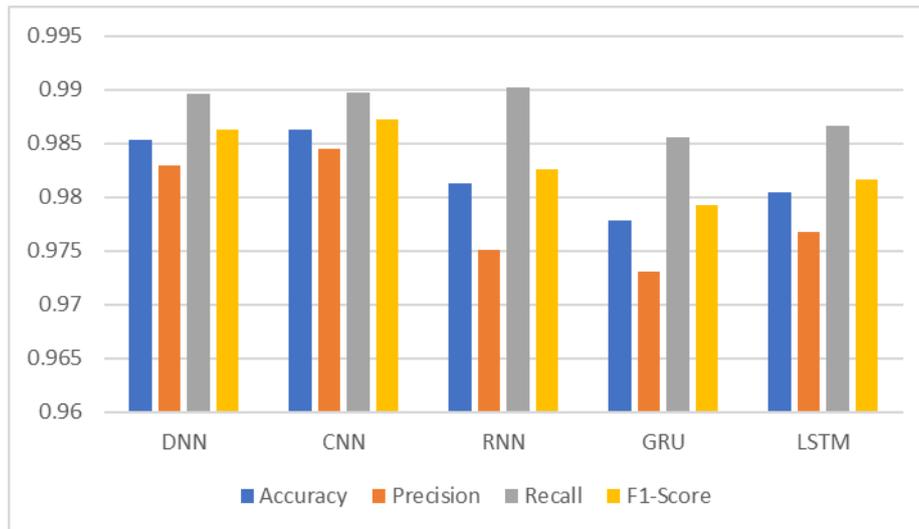

Figure 7.  Evaluation Metrics of Deep learning classifiers

The goal of classifiers during their evaluation on the confusion matrix is to obtain the highest value of the measures (TP, TN) and reduce the value of (FP, FN) as much as possible. The CNN classifier has the highest value (TP) and the lowest value (FP) as shown in Table 9. The RNN classifier has a higher value (TN) than all other classifiers. The DNN classifier got results in (TP, TN) higher than the classifier LSTM, GRU and also higher RNN in the parameter (TP). The rest of classifiers like LSTM achieve better results in parameter (TP, TN, FP, FN) than GRU classifier, the results of these algorithms are shown in Table 4.

Table 4.  Evaluation Metrics Classifiers.

| DL-classifiers | Confusion Matrix-Parameters | | | |
|---|---|---|---|---|
| | TP | TN | FP | FN |
| DNN | 14433 | 16601 | 287 | 173 |
| CNN | 14460 | 16604 | 260 | 170 |
| RNN | 14262 | 16611 | 424 | 163 |
| LSTM | 14326 | 16550 | 394 | 224 |
| GRU | 14262 | 16534 | 224 | 240 |

Another important metric, such as ROC (Receiver Operating Curve), by which the results of deep learning classifiers are evaluated, are shown in Table 10. The results of the algorithms DNN, CNN are similar, so the result of the classifier is (0.998). The algorithms RNN and LSTM also obtained the same results (0.997), the GRU algorithm obtained (0.996) as in Table 5.

Table 5. ROC Metrics.

| Algorithm | ROC |
|-----------|-------|
| DNN | 0.998 |
| CNN | 0.998 |
| RNN | 0.997 |
| LSTM | 0.997 |
| GRU | 0.996 |

## 7. STUDY COMPARATIVE

In this section, we will discuss and compare our approach to results with another related study.

In [8], the author had a detection accuracy of 75.75% on the binary classifier. Similarly, the same author in [9] achieved a detection accuracy result of 82.02% using the hybrid approach from the deep learning classifier. The author in [13] achieved a detection accuracy of 93.72% using the LSTM classifier. In [14], more than one machine learning classifier was used and good results were obtained. Compared with previous results and methods, our approach provides an accurate description of the methods used to process the data set, and it uses multiple classifiers to measure the impact of the same method used for the results, in addition, our approach is also based on the extraction of features that affect the results, leading to the performance of the training and high detection process. Our approach to evaluating results also relies on a variety of different metrics. A comparison of the studies is presented in Table 6.

Table 6. Accuracy Result Comparison with another Study Related.

| Ref. | Method | Dataset | Accuracy |
|------|--------|---------|----------|
| [8] | DNN | NSL-KDD | 75.75 % |
| [9] | GRU-RNN | NSL-KDD | 82.02 % |
| [13] | LSTM | CSIC 2010 | 93.72 % |
| [14] | LR, | DS2OS traffic | 98.3 % |
|  | SVM, | traces | 98.2 % |
|  | DT, |  | 99.4 % |
|  | RF, |  | 99.4 % |
|  | ANN |  | 99.4 % |
| Our Method (NIDS-DL) | CNN | NSL-KDD | 98.63 % |
|  | DNN |  | 98.53 % |
|  | RNN |  | 98.13 % |
|  | LSTM |  | 98.04 % |
|  | GRU |  | 97.78 % |

## 8. CONCLUSION

In this paper, more than one type of deep learning algorithm is used and applied to detect abnormality in NIDS. The approach was evaluated on different metrics and the approach achieved high and reliable results. One of the most contributions of this work is using the feature selection method to train the classifiers on most feature correlations and avoid miss led during training to reach the best result. Our approach focused on binary classification using deep learning algorithms. The results of the algorithms are compared with each other, the results of some classifiers are close, and the CNN classifier achieved the highest results. The use of deep learning demonstrated the possibility and superiority when applied in the binary classification of network intrusion detection systems. Since the proposed approach harvest high results, future work will be to evaluate the results of classifiers on more than one type of dataset and compare the results. A hybrid approach of deep learning algorithms can also be used as a future work, and its results compared with our approach. These approaches can also be used to detect a specific type of attack, such as (DOS) attacks also we apply this approach inside SDN environment.

## REFERENCES

[1] "Software-Defined Networking (SDN) Definition - Open Networking Foundation." https://opennetworking.org/sdn-definition/ (accessed Apr. 25, 2022).

[2] McKeownNick *et al.*, "OpenFlow," *ACM SIGCOMM Comput. Commun. Rev.*, vol. 38, no. 2, pp. 69–74, Mar. 2008, doi: 10.1145/1355734.1355746.

[3] S. Sangeetha, B. Gayathri Devi, R. Ramya, M. K. Dharani, and P. Sathya, "Signature Based Semantic Intrusion Detection System on Cloud," *Adv. Intell. Syst. Comput.*, vol. 339, pp. 657–666, 2015, doi: 10.1007/978-81-322-2250-7_66.

[4] S. K. Dey and M. M. Rahman, "Effects of Machine Learning Approach in Flow-Based Anomaly Detection on Software-Defined Networking," *Symmetry 2020, Vol. 12, Page 7*, vol. 12, no. 1, p. 7, Dec. 2019, doi: 10.3390/SYM12010007.

[5] R. Sommer and V. Paxson, "Outside the closed world: On using machine learning for network intrusion detection," *Proc. - IEEE Symp. Secur. Priv.*, pp. 305–316, 2010, doi: 10.1109/SP.2010.25.

[6] M. Tavallaee, E. Bagheri, W. Lu, and A. A. Ghorbani, "A detailed analysis of the KDD CUP 99 data set," *IEEE Symp. Comput. Intell. Secur. Def. Appl. CISDA 2009*, Dec. 2009, doi: 10.1109/CISDA.2009.5356528.

[7] S. Revathi and A. Malathi, "A Detailed Analysis on NSL-KDD Dataset Using Various Machine Learning Techniques for Intrusion Detection," *Int. J. Eng. Res. Technol.*, 2013.

[8] T. A. Tang, L. Mhamdi, D. McLernon, S. A. R. Zaidi, and M. Ghogho, "Deep learning approach for Network Intrusion Detection in Software Defined Networking," *Proc. - 2016 Int. Conf. Wirel. Networks Mob. Commun. WINCOM 2016 Green Commun. Netw.*, pp. 258–263, Dec. 2016, doi: 10.1109/WINCOM.2016.7777224.

[9] T. A. Tang, L. Mhamdi, D. McLernon, S. A. R. Zaidi, and M. Ghogho, "Deep Recurrent Neural Network for Intrusion Detection in SDN-based Networks," *2018 4th IEEE Conf. Netw. Softwarization Work. NetSoft 2018*, pp. 462–469, Sep. 2018, doi: 10.1109/NETSOFT.2018.8460090.

[10] I. I. Kurochkin and S. S. Volkov, "Using GRU based deep neural network for intrusion detection in software-defined networks," *IOP Conf. Ser. Mater. Sci. Eng.*, vol. 927, no. 1, Sep. 2020, doi: 10.1088/1757-899X/927/1/012035.

[11] "NSL-KDD | Datasets | Research | Canadian Institute for Cybersecurity | UNB." https://www.unb.ca/cic/datasets/nsl.html (accessed Apr. 25, 2022).

[12] "KDD Cup 1999 Data." http://kdd.ics.uci.edu/databases/kddcup99/kddcup99.html (accessed Apr. 25, 2022).

[13] S. Althubiti, W. Nick, J. Mason, X. Yuan, and A. Esterline, "Applying Long Short-Term Memory Recurrent Neural Network for Intrusion Detection," *Conf. Proc. - IEEE SOUTHEASTCON*, vol. 2018-April, Oct. 2018, doi: 10.1109/SECON.2018.8478898.

[14] M. Hasan, M. M. Islam, M. I. I. Zarif, and M. M. A. Hashem, "Attack and anomaly detection in IoT sensors in IoT sites using machine learning approaches," *Internet of Things*, vol. 7, p. 100059, Sep. 2019, doi: 10.1016/J.IOT.2019.100059.

## AUTHORS

**Mhmood Radhi Hadi** is a master's student of Computer engineering at Karabük University, Turkey. Before joining Karabük university in 2020, he was getting BSc degree in 2019 from Networking engineering at Iraqi University, Iraq. Before to university engagement he was working as a graphic designer, and through the BSc journey he was a working as internship and training Network simulation in Cisco academy also through the MSc period he was getting internship in Curve AI company as A Machine Learning. His research interests are Network security using AI/ML/DL, Software-defined Network, wireless and communication, Intelligent system.

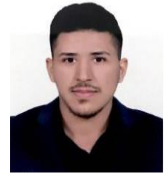

**Adnan Saher Mohammed** received his B.Sc. degree in computer engineering technology in 1999 from Northern technical university, Mosul, Iraq. In 2012 obtained an M.Sc. degree in communication and computer network engineering from UNITEN University, Kuala Lampur, Malaysia, and in 2017 received Ph.D. degree from Ankara Yildirim Beyazit University, Ankara, Turkey. He is currently assistant professor at Karabük university, Turkey. His research interests include computer networks, Algorithms and Artificial Intelligence.

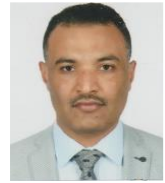